\documentstyle[11pt]{article}
\newcommand{\ind}[1]{}
\newcommand{\zpsfig}[1]{}
\newtheorem{prop}{Relation}
\newcommand{\secb}{\subsection}
\newcommand{\seca}{\section}
\newcommand{\appenc}{appendix~B}
\newcommand{\eg}{{\it e.g.}}
\newcommand{\cs}{charged scalar}
\newcommand{\css}{charged scalars}
\newcommand{\nfcl}{\mbox{$95\%~{\rm CL}$}}
\newcommand{\ncl}{\mbox{$90\%~{\rm CL}$}}
\newcommand{\brsm}{\mbox{${\rm BR}^{SM}$}}
\newcommand{\brmh}{\mbox{${\rm BR}^{MHDM}$}}
\newcommand{\BR}{branching ratio}
\newcommand{\rarrow}{\rightarrow}
\newcommand{\tgb}{\mbox{$\tan\beta$}}
\newcommand{\vtd}{\mbox{$|V_{td}|$}}
\newcommand{\lef}{\left}
\newcommand{\rit}{\right}

\newcommand{\tab}{~~~~~~}
\newcommand{\Gev}{{\rm GeV}}
\newcommand{\Mev}{{\rm MeV}}
\newcommand{\cpv}{CP violation}
\newcommand{\chbd}{charged scalar box diagrams}
\newcommand{\smbd}{SM box diagrams}
\newcommand{\chpd}{charged scalar penguin diagrams}

\newcommand{\moda}{{\it Model I\/}}
\newcommand{\modb}{{\it Model II\/}}
\newcommand{\modc}{{\it Model III\/}}
\newcommand{\beq}{\begin{equation}}
\newcommand{\eeq}{\end{equation}}
\newcommand{\bear}{\begin{array}}
\newcommand{\eear}{\end{array}}
\newcommand{\beqa}{\begin{eqnarray}}
\newcommand{\eeqa}{\end{eqnarray}}
\newcommand{\ben}{\begin{enumerate}}
\newcommand{\een}{\end{enumerate}}
\newcommand{\bei}{\begin{itemize}}
\newcommand{\eei}{\end{itemize}}
\newcommand{\bef}{\begin{figure}}
\newcommand{\eef}{\end{figure}}
\newcommand{\xyzi}{$X_i$, $Y_i$ and $Z_i$}
\newcommand{\xyz}{$X$, $Y$ and $Z$}
\newcommand{\xy}{\mbox{$XY^*$}}
\newcommand{\imxy}{\mbox{Im($XY^*$)}}
\newcommand{\imxz}{\mbox{Im($XZ^*$)}}
\newcommand{\yx}{\mbox{$YX^*$}}
\newcommand{\xz}{\mbox{$XZ^*$}}
\newcommand{\yz}{\mbox{$YZ^*$}}
\newcommand{\delkm}{\mbox{$\delta_{\rm CKM}$}}
\newcommand{\cont}{\nonumber \\}
\newcommand{\ihh}[1]{I_{HH}^{#1}}
\newcommand{\ihw}[1]{I_{HW}^{#1}}

\newcommand{\mws}{\mbox{$m_W^2$}}

\newcommand{\mzs}{\mbox{$m_Z^2$}}
\newcommand{\mh}{\mbox{$m_H$}}
\newcommand{\mhs}{\mbox{$m_H^2$}}
\newcommand{\mtau}{\mbox{$m_{\tau}$}}
\newcommand{\mmu}{\mbox{$m_{\mu}$}}
\newcommand{\mt}{\mbox{$m_t$}}
\newcommand{\mts}{\mbox{$m_t^2$}}
\newcommand{\mb}{\mbox{$m_b$}}
\newcommand{\ys}{{|Y|}^2}
\newcommand{\yf}{{|Y|}^4}
\newcommand{\xs}{{|X|}^2}
\newcommand{\xf}{{|X|}^4}
\newcommand{\zs}{{|Z|}^2}
\newcommand{\xys}{{(XY^*)}^2}

\newcommand{\vll}{V_{LL}}
\newcommand{\sll}{S_{LL}}

\newcommand{\viavll}{\langle\bar{B}|
                     [\bar{d}\gamma^{\nu}(1-\gamma_5)b]
                     [\bar{d}\gamma_{\nu}(1-\gamma_5)b]
                     |B\rangle}
\newcommand{\viasll}{\langle\bar{B}|
                     [\bar{d}(1-\gamma_5)b]
                     [\bar{d}(1-\gamma_5)b]
                     |B\rangle}
\newcommand{\kk}{\mbox{$K-\bar{K}$}}
\newcommand{\dd}{\mbox{$D-\bar{D}$}}
\newcommand{\bb}{\mbox{$B-\bar{B}$}}
\newcommand{\bsbs}{\mbox{$B_s-\bar{B}_s$}}
\newcommand{\bsg}{\mbox{$b \rightarrow s \gamma$}}

\newcommand{\kpinunu}{\mbox{$K^+ \rightarrow \pi^+ \nu \bar{\nu}$}}
\newcommand{\bknunu}{\mbox{$B \rightarrow X_s \nu \bar{\nu}$}}
\newcommand{\bxtaunu}{\mbox{$B \rightarrow X \tau \nu_{\tau}$}}

\newcommand{\taue}{\mbox{$\tau \rightarrow e \nu_\tau \bar{\nu}_e$}}
\newcommand{\taumu}{\mbox{$\tau \rightarrow \mu\nu_{\tau}\bar{\nu}_\mu$}}

\newcommand{\bll}{\mbox{$B \rightarrow {\ell}^+ {\ell}^-$}}

\newcommand{\blnu}{\mbox{$B \rightarrow {\ell} \nu$}}

\newcommand{\bxll}{\mbox{$B \rightarrow X_s {\ell}^+ {\ell}^-$}}
\newcommand{\btautau}{\mbox{$B \rightarrow {\tau}^+ {\tau}^-$}}
\newcommand{\bstautau}{\mbox{$B_s \rightarrow {\tau}^+ {\tau}^-$}}
\newcommand{\bxtautau}{\mbox{$B \rightarrow X_s {\tau}^+ {\tau}^-$}}
\newcommand{\kmumu}{\mbox{$K_L \rightarrow {\mu}^+ {\mu}^-$}}
\newcommand{\bmunu}{\mbox{$B \rightarrow {\mu} {\nu}$}}
\newcommand{\btaunu}{\mbox{$B \rightarrow {\tau} {\nu}$}}
\newcommand{\zbb}{\mbox{$Z \rightarrow b \bar{b}$}}
\newcommand{\ztautau}{\mbox{$Z \rightarrow \tau \bar{\tau}$}}
\newcommand{\epb}{\mbox{$\epsilon_b$}}
\newcommand{\db}{\mbox{$\nabla_b$}}
\newcommand{\dbsm}{\mbox{$\nabla_b^{SM}$}}
\newcommand{\dbh}{\mbox{$\nabla_b^H$}}

\newcommand{\tnu}{\mbox{$\tau \nu$}}
\newcommand{\scb}{\mbox{$cs$}}
\newcommand{\bc}{\mbox{$cb$}}
\newcommand{\sws}{\mbox{$\sin^2 \theta_W$}}
\begin{document}
\baselineskip 0.23in
\begin{titlepage}
\begin{center}
\vspace*{1cm}
\begin{flushright}
WIS-94/3/Jan-PH \\
January 1994
\end{flushright}
\vspace{25mm}
{\LARGE Phenomenology of models with more than two Higgs doublets}
\\ \vspace{15mm}
{\Large Yuval Grossman}
\\ \vspace{5mm}
{Department of Particles Physics \\
The Weizmann Institute of Science \\
Rehovot 76100, ISRAEL}
\\ \vspace{10mm}
\end{center}
\begin{abstract}
We study the most general Multi-Higgs-Doublet Model (MHDM)
with Natural Flavor Conservation (NFC).
The couplings of a charged scalar $H_i^{\pm}$ to
up quarks, down quarks and charged leptons depend on three
new complex parameters, $X_i$, $Y_i$ and $Z_i$, respectively.
We prove relations among these parameters.
We carry out a comprehensive analysis of phenomenological
constraints on the couplings of the lightest charged scalar:
$X$, $Y$ and $Z$.
We find that the general MHDM may differ significantly
from its minimal version, the Two-Higgs-Doublet Model (2HDM).
\end{abstract}

\end{titlepage}

\seca{Introduction}
The Higgs sector of the Standard Model (SM), consisting of a single
Higgs doublet, has not yet been experimentally tested.
The possibility of an extended Higgs sector is definitely
consistent with experimental data.
The simplest extensions are models with several
Higgs doublets, implying
the existence of \css. The simplest of these is
the Two-Higgs-Doublet Model
(2HDM), which has received substantial
attention in the literature~\cite{hunter}.
However, very little is found in the literature on models with
more than two Higgs doublets and, in particular, there
exists no comprehensive analysis of these models.
It is the purpose of this work to study
the general Multi-Higgs-Doublet Model (MHDM), and perform a
comprehensive analysis of the charged Higgs sector of the model.
In the lack of direct experimental
knowledge, we study
indirect effects of \css\ in order to place constraints on this sector.
In particular, we consider \cs\ contributions to low-energy
processes: neutral meson mixing, rare $K$ and $B$ decays,
$Z$ boson interactions and CP violating effects.
The results from this combined analysis are used to
find the region in parameter space which is consistent with all data.
We then discuss the differences between the MHDM and the 2HDM.
Throughout this work we assume that there are no significant
cancellations with other types of new physics.

Our work is organized as follows. In chapter~2 we describe the MHDM
and prove general relations between the model parameters.
In chapter~3 we discuss the relevant processes,
and list the formulae and numerical data used in our analysis.
In chapter~4 we combine all the constraints
and find the allowed range of the
model parameters. In chapter~5 we discuss the differences
between the MHDM and the 2HDM.
Finally, chapter~6 contains a summary of our work.

\seca{The Model}
\secb{Introduction}
There are basically two major constraints on any
extension of the Higgs sector of the SM.
First, it is an experimental fact
that $\rho=\mws/(\mzs\cos^2\theta_W)$ is very close to 1~\cite{pdg}.
It is known~\cite{hunter}
that in a model with only Higgs doublets, the tree-level
value of $\rho=1$ is automatic without adjustment of any parameters in
the model. Second, there are strong experimental limits on
flavor-changing neutral currents (FCNC's). In the SM, tree-level FCNC's
are absent because
fermion mass matrices and Higgs-fermion couplings are
simultaneously diagonalized.
In general, this ceases to be true in a model with
a non-minimal Higgs sector. An elegant
way to avoid this problem is based on a theorem by Glashow and
Weinberg~\cite{glwein} called Natural Flavor Conservation (NFC):
tree-level FCNC's mediated by Higgs bosons will be
absent if all fermions of a given electric charge couple to no more than
one Higgs doublet. This can be achieved by imposing extra symmetries.
If we adopt this mechanism, the Higgs
couplings to fermions are constrained but not unique. There are five
possibilities to couple the Higgs doublets to the known three types of
massive fermions (up-type quarks, down-type quarks and charged leptons.).
Summary of these possibilities is given in
Table~\ref{tb:models}. For \modc\ at least three Higgs
doublets are needed, for \moda\ one is enough,
while for the other models two are sufficient. It does, of course,
make a difference if the number of
Higgs doublets is larger than the minimal one.
\begin{table}
\begin{center}
\begin{tabular}{l|ccccc|}
  \multicolumn{1}{c}{} & \multicolumn{5}{c}{\it Model} \\ \cline{2-6}
  & \it{I} & \it{I'} & \it{II} & \it{II'} & \it{III} \\ \cline{2-6}
d (down-type quarks) & 1 & 1 & 1 & 1 & 1 \\
u (up-type quarks)   & 1 & 1 & 2 & 2 & 2 \\
e (charged leptons)  & 1 & 2 & 2 & 1 & 3 \\ \cline{2-6}
\end{tabular}
\end{center}
\caption[tbmodels]
{Summary of all the possibilities of MHDM. The numbers
in the table show which
Higgs doublet couples to which fermion type.}
\label{tb:models}
\end{table}

For a general MHDM the Yukawa interactions are given by
\beq
-{\cal L}_Y={\overline{Q}}_{L_i} F_{ij}^D \Phi_d D_{R_j} +
          {\overline{Q}}_{L_i} F_{ij}^U \tilde{\Phi}_u U_{R_j} +
          {\overline{L}}_{L_i} F_{ij}^E \Phi_e E_{R_j} + {\rm h.c.}
\eeq
Where left-handed quark doublets are denoted by $Q_{L_i}$,
and left-handed lepton doublets by $L_{L_i}$.
Right-handed quark
singlets are denoted by $D_{R_i}$ and $U_{R_i}$, and right handed
charged lepton singlets by $E_{R_i}$.
The sub-index $i$ is a generation index ($i=1,2,3$).
Higgs doublets are denoted by $\Phi_j$ (where $j$ runs from 1 to $n$,
the number of Higgs doublets), and $\tilde{\Phi}_j=i\sigma_2\Phi_j^*$.
Sub-indices $d$, $u$ and $e$ denote the Higgs doublet that
couples to the down-type quarks, up-type quarks and charged
leptons, respectively.
$F^U$ and $F^D$ are general $3\times3$ Yukawa matrices
and we can choose a basis where one
of them is real and diagonal.
Since we consider massless neutrinos,
$F^E$ can be chosen real and diagonal.
In a general MHDM, with $n$ Higgs doublets,
there are $2n$ charged and $2n$ neutral scalar
fields. After spontaneous symmetry breaking two charged fields and
one neutral field become the would-be
Goldstone boson ``eaten" by the $W^\pm$ and
the $Z$ in order to acquire their masses, and only
$2(n-1)$ charged and $2n-1$ neutral physical scalars are left.
The Yukawa interaction of the physical
\css\ with fermion mass eigenstates is given by~\cite{tye}
\beq
{\cal L}_Y^{\pm}=(2\sqrt{2}G_F)^{1/2} \sum_{i=2}^{n}
(X_i\overline{U}_L V M_D D_R +
 Y_i\overline{U}_R M_U V D_L +
 Z_i\overline{N}_L M_E E_R ) H_i^+ + {\rm h.c.}
\eeq
$M_D$, $M_U$ and $M_E$ are the diagonal mass matrices of down-type
quarks, up-type quarks and charged leptons, respectively.
$H^+_i$ denote the positively charged physical scalars.
Left-handed neutrino fields are denoted by $N_L$, and the
CKM matrix by $V$.
\xyzi\ are complex coupling constants that arise from the
mixing matrix for charged scalars.
An important feature of the \cs\ -- fermion
interactions is that they are
proportional to the fermion masses. Hence,
effects are stronger in processes involving heavy fermions,
real or virtual.

\secb{Relations between \xyzi}
\xyzi\ arise from the mixing matrix $U$,
which rotates the \cs\ interaction eigenstates into the mass
eigenstates~\cite{tye}
\beq   \label{eq:diag}
\left( \bear{c} G^+ \\ H_2^+ \\ \vdots\\ H_n^+ \eear \right) =
U \left(\bear{c}\Phi_1^+ \\ \Phi_2^+ \\ \vdots \\ \Phi_n^+ \eear \right),
\eeq
where $G^+$ is the would-be
Goldstone boson and $H^+_i$ are the physical charged scalars.
$U$ is an $n\times n$ unitary matrix that can be parameterized
as the generalized CKM matrix with
$n(n-1)/2$ angles and $(n-1)(n-2)/2$ phases~\cite{tye}.
Since $G^+$ is the Goldstone boson that gives the $W^+$ its mass
in the same way as in the SM, we obtain
\beq
G^+=\frac{1}{v}\sum_{i=1}^n v_i \Phi_i^+,
\eeq
where $v_i\equiv <\Phi_i>$ and
$v\cong 246~\Gev$ is the SM vacuum expectation value (vev).
Using eq.~(\ref{eq:diag}) we get
\beq  \label{eq:defvi}  \label{eq:defv}
v_i=U_{1i}v,  \tab
\sum_{i=1}^n v_i^2=v^2.
\eeq
We define
\beq   \label{eq:defxyz}
X_i=\frac{U_{di}}{U_{d1}}, \tab
Y_i=-\frac{U_{ui}}{U_{u1}}, \tab
Z_i=\frac{U_{ei}}{U_{e1}}.
\eeq
Using these definitions and
the unitarity of $U$, we obtain the following relations:
\begin{prop} \label{th:1}
\beqa \label{eq:th1}
\sum_{i=2}^n X_iY_i^*=1& & \mbox{{\rm(for $\Phi_d\not=\Phi_u$)}},  \\
\sum_{i=2}^n X_iZ_i^*=-1& & \mbox{{\rm(for $\Phi_d\not=\Phi_e$)}}, \nonumber \\
\sum_{i=2}^n Y_iZ_i^*=1& & \mbox{{\rm(for $\Phi_u\not=\Phi_e$)}}.  \nonumber
\eeqa
\end{prop}
\begin{prop} \label{th:2}
\beq  \label{eq:th2}
\sum_{i=2}^n {|X_i|}^2= \frac{v^2}{v_d^2}-1, \tab
\sum_{i=2}^n {|Y_i|}^2= \frac{v^2}{v_u^2}-1, \tab
\sum_{i=2}^n {|Z_i|}^2= \frac{v^2}{v_e^2}-1.
\eeq
\end{prop}
\begin{prop} \label{th:3}
{\rm $X_i$, $Y_j$ and $Z_k$ $(i,j,k\in 2..n)$
cannot be simultaneously large
(more than $\sqrt{n-1}$), unless there is at
least one Higgs doublet that does not couple to fermions.}
\end{prop}
These relations are proved in appendix A.

Special cases of our results are known for 2HDM and 3HDM.
In 2HDM there is only one \cs, so we
drop the sub-indices of \xyzi. For \modb~\cite{hunter}
\beq
Z=X=\tan\beta, \tab
Y=\cot\beta, \tab
\tan\beta=\frac{v_u}{v_d},
\eeq
and the three relations are clearly fulfilled.
For 3HDM~\cite{bbg}
\beq
X_1Y_1^*=1-X_2Y_2^*, \tab
Y_1^2+Y_2^2=\frac{v_d^2+v_e^2}{v_u^2}, \tab
X_1^2+X_2^2=\frac{v_u^2+v_e^2}{v_d^2},
\eeq
which are examples of relations~\ref{th:1} and \ref{th:2}.

A few points are in order:
\ben
\item{An important relation between $X$ and $Y$ in 2HDM (\modb), namely
$\xy=1$, is {\em not} retained in the framework of a general MHDM.
Consequently, processes involving this combination can be enhanced
or suppressed in MHDM compared to 2HDM. A similar statement holds for
the combinations \xz\ and \yz.}
\item{The most general features of
MHDM can be realized in 4HDM. In 3HDM (\modc) the couplings
cannot be simultaneously large. This constraint is removed
in 4HDM. However, in most of the processes that we study, not all
couplings are involved, and 3HDM is general
enough.}
\item{From relation~\ref{th:1} we see that if all
\css\ were degenerate, the model practically reduces to 2HDM,
namely, the relations $\xy=1$, $\xz=-1$ and $\yz=1$ hold.
Consequently, in order to exploit
the most general MHDM,
we assume that one of the \css\
is much lighter than the others, and
that all the heavy \css\
effectively decouple from the fermions.
Then, we consider in the calculations the single light \cs,
and drop the sub-indices of $H_i$, \xyzi.}
\een

\secb{\cpv}
Within the framework of MHDM,
\cpv\ could arise in \cs\ exchange if there are at least
three Higgs doublets~\cite{wein}.
Then, there are two mechanisms which give rise to \cpv\
(for a review see~\cite{bbg}):
\ben
\item{Spontaneous \cpv\ (SCPV).
CP symmetry is enforced at the Lagrangian level, but
broken through complex vev's.
The requirement of SCPV forces $\delkm=0$~\cite{branco}.
In this case, CP noninvariance arises solely from
\cs\ exchange. However, experimental data
exclude this possibility~\cite{pok2,yuval}.}
\item{Explicit \cpv.
CP is broken by complex Yukawa couplings and possibly by
complex vev's\footnote{
Actually, CP can also be broken through
complex scalar couplings. We will not discuss this possibility.}.
In such a case, \cpv\ can arise from both
\cs\ exchange and $W^\pm$ exchange.}
\een
In both cases, \cpv\ in the Higgs sector is manifest in
phases that appear
in the combinations \xy, \xz\ and \yz.
We emphasize that the requirement that the \css\
are nondegenerate is mandatory for \cpv\ from the
Higgs sector. If the \css\ were degenerate, \xy\ is
replaced by $\sum_{i=2}^n X_iY_i^*=1$, which is real.

\secb{Summary}
\bef
\ind{digrule}
\caption[fifeynman]
{The Feynman rules for the \cs\ - fermions interaction within MHDM.}
\label{fi:feynman}
\eef
The model that we study is the SM with extended Higgs sector. Its
Higgs sector contains more than two Higgs doublets with NFC.
We focus on the \css, and assume that
all but the lightest of the \css\
effectively decouple from fermions.
The couplings of the lightest \cs\ to fermions are given by:
\beq
{\cal L}_Y^+=(2\sqrt{2}G_F)^{1/2}
 (X\overline{U}_L V M_D D_R +
  Y\overline{U}_R M_U V D_L +
  Z\overline{N}_L M_E E_R ) H^+ + {\rm h.c.}
\eeq
This interaction yields the Feynman rules given in fig~\ref{fi:feynman}.
In this model there are several new parameters: The mass of
the lightest \cs: \mh, and the coupling constants: \xyz.
These couplings are {\em arbitrary} complex numbers.
It is the subject of the coming
chapters to find the experimentally allowed ranges for these parameters.

\seca{Phenomenology}
In this chapter we discuss various
processes where a virtual \cs\
may contribute significantly. From the experimental results
for these processes, we are able to constrain the parameters of
the model.
Wherever possible we use the experimental data at the \nfcl.
For theoretical uncertainties we use estimates and cannot
assign a confidence level to the errors.
Before discussing the processes, we list values of
quark sector parameters that we use,
and give constraints on the \cs\ parameters that are obtained
from direct measurements and perturbativity.

\secb{Values of quark parameters}
For quark masses we use~\cite{qmass}
\beq \label{qmass}
m_u=5~\Mev, \tab
m_d=9~\Mev, \tab
m_s=180~\Mev, \tab
m_c=1.4-1.5~\Gev, \tab
m_b=4.6-5~\Gev.
\eeq
For the top quark the situation is different.
First, it is the only quark that has
not been observed. Thus, there is large uncertainty in its mass.
Second, since it is the heaviest quark,
its coupling to the \cs\ is the largest.
The direct search
for the top at CDF gives $\mt>113$ GeV~\cite{cdf}, while the
electroweak precision measurements at LEP give $\mt<185$ GeV~\cite{lep}.
Within the framework of MHDM both bounds are relaxed.
If $m_H < m_t$, the top may escape detection at CDF~\cite{barger2}.
Additional loops that include the extra scalars may affect the
electroweak precision measurements~\cite{guth}.
Then, the bounds on $m_t$ are related to the MHDM parameters.
Thus, whenever we use some values of the MHDM parameters we use
only values of $m_t$ that are consistent with them.

For the values of the CKM matrix elements we use~\cite{yckm,nirsarid,ship}
\beqa
|V_{ud}|\cong |V_{cs}|\cong |V_{tb}| &\cong& 1,   \\
|V_{cd}|\cong |V_{us}|&\cong& 0.22,      \cont
|V_{ts}|\cong |V_{cb}|&=&0.040 \pm 0.007, \cont
|V_{ub}/V_{cb}|&=&0.08 \pm 0.03.    \nonumber
\eeqa
Unitarity constraints on \vtd\ give
\beq \label{eq:nvtd}
|V_{td}|=0.004-0.016.
\eeq
The allowed ranges for \vtd\ and
\delkm\ are different in the SM and in
MHDM~\cite{buras}. Furthermore, they are related and depend on \mt.

For the masses of the physical particles we use
the values quoted by
the Particles Data Group (PDG)~\cite{pdg}.

\secb{Constraints on scalar parameters}
For the mass of the lightest \cs\ we use~\cite{ron}
\beq
\mh > 43 ~\Gev.
\eeq
There is no
upper bound on \mh: The SM Higgs boson is bounded by unitarity
to be lighter than about
$1$ TeV, but there is no similar unitarity bound on the \cs\
of MHDM (for a discussion of this point see~\cite{hunter}).

We can get perturbativity bounds on $|X|$, $|Y|$ and $|Z|$:
We restrict ourselves to the range where
perturbation expansion is valid. Following~\cite{barger} we find
\beqa
|Y|&<&\frac{600~\Gev}{\mt}<7, \label{eq:pert} \\
|X|&<&\frac{600~\Gev}{\mb}\approx 130, \cont
|Z|&<&\frac{600~\Gev}{\mtau}\approx 340.  \nonumber
\eeqa
There is no lower bound on $|X|$, $|Y|$ and $|Z|$.
When all of them vanish, the model practically reduces to the SM.

\secb{\bb\ mixing}
The mass difference in the neutral $B$ meson system ($\Delta m_B$)
has been measured and it fits
the SM predictions \cite{fran,yckm,ycpv}. However, there
are large uncertainties in the calculation (mainly in the
values of \mt, $V_{td}$ and $f_B$)
that leave a lot of room for non-SM contributions.
The MHDM extra contributions to the mixing come from box diagrams
with intermediate \css\ (see fig.~\ref{fi:bbbox}).
\bef
\ind{dig1}
\caption[fibbbox]
{The diagrams that contribute to \bb\ mixing. (a): The SM diagram.
(b): The two-Higgs diagram. (c): The mixed diagrams.
The diagrams where the intermediate bosons and fermions are exchanged
are not shown.}
\label{fi:bbbox}
\eef

In the calculation
we neglect external momenta but keep all quark masses.
We use the Vacuum Insertion Approximation (VIA) which is
a good approximation in this case~\cite{sach}.
For intermediate quark only the top quark is important
\beq  \label{eq:monetwo}
M_{12}(B_q)=\frac{G_F^2}{64{\pi}^2} \mws \eta_B (V_{tq}^*V_{tb})^2
[ I_{WW}+\ihh{}+\ihw{} ] \tab q=d,s.
\eeq
The SM contribution (fig.~\ref{fi:bbbox}(a)) is $I_{WW}$\cite{fran},
while box diagrams with two \cs\
propagators (fig.~\ref{fi:bbbox}(b))
or one \cs\ and one $W$ propagator (fig.~\ref{fi:bbbox}(c))
give $\ihh{}$ and $\ihw{}$, respectively.
$\eta_B\sim0.55$ is a QCD correction factor~\cite{jamin}.
The QCD corrections to the MHDM diagrams have been calculated
in the leading logarithmic approximation~\cite{bbhls}, and
only for the terms with the same Dirac structure as the SM diagrams.
However, following~\cite{barger}
we apply the same QCD correction factor
both to the SM and the MHDM contributions.
In what follows we present only the most important terms:
\beqa \label{eq:iwwhh}
I_{WW}&=&x_t 4I_0(x_t) \vll, \\
\ihh{}&=&x_t [ y_t \yf I_1(y_t) \vll +
 y_t y_b \xys 4I_2(y_t) \sll  +
 y_b y_q \xf I_1(y_t) \vll ],  \cont
\ihw{}&=&x_t [ x_t \ys (2I_3(x_t,x_H)-8I_4(x_t,x_H)) \vll +
 2x_t x_b (\yx) I_4(x_t,x_H) \sll ],   \nonumber
\eeqa
where $x_q \equiv m_q^2/\mws$ and $y_q \equiv m_q^2/\mhs$.
The matrix elements, $\vll$ and $\sll$, and
the loop integrals, $I_i$, are given in \appenc.
Setting $m_b$ and $m_d$ to zero we get the known results~\cite{wise,gil},
while for the case where only $m_d$ is set to zero
our results agree with those of \cite{pok2}.
We use $\Delta m_B = 2 |M_{12}|$
and~\cite{pdg,mati,ship}
\beq \label{eq:nbb}
x_d \equiv \Delta m_B / \Gamma_B=0.67 \pm 0.11, \tab
f_B=190 \pm 50 ~\Mev, \tab
\tau_B=1.49 \pm 0.03 {\rm ~ps}.
\eeq

We  conclude  that large contribution to \bb\ mixing
may arise for large $Y$ or large \xy.
For the $B$ system the contribution from the term
proportional to $\xf$ is small,
even for $X$ as large as the perturbativity bound.
However, for the $B_s$ system its contribution can be large.

\secb{\kk\ mixing}
There are two well measured
quantities related to \kk\ mixing:
the mass difference $\Delta m_K$
and the CP violating observable $\epsilon$
(for a review see \cite{yckm}).
Taking into account the bounds from \bb\ mixing, we find
that charged scalar mediated diagrams
may give only a small contribution to $\Delta m_K$~\cite{hunter}.
The situation is different for $\epsilon$~\cite{ycpv}
\beq
|\epsilon|=\frac{{\rm Im}(M_{12})}{\sqrt{2}\Delta M}.
\eeq
$\epsilon$ is a very accurately measured quantity,
but there are large uncertainties in the SM
calculation (mainly from \mt, $V_{td}$ and $B_K$).
In MHDM there are two new sources that can contribute
to $\epsilon$: The imaginary part of the \chbd, and
enhanced long-distance contributions.
The short-distance diagrams are similar to those of
\bb\ mixing (fig.~\ref{fi:bbbox}) with the replacement
$b \rightarrow s$.
For the \smbd\ we consider
both charm and top intermediate quarks,
while for the \chbd\
only the top quark is important. We keep only the terms that are
proportional to the internal quark masses:
\beq
|\epsilon_{SD}|=\frac{1}{\sqrt{2}\Delta M}
\frac{G_F^2}{48{\pi}^2}
\mws m_K f_K^2 B_K~{\rm Im}[I^c + I^t + I^{ct}] \label{eq:epsilon},
\eeq
with
\beqa
I^c&=&\eta_1 x_c \lambda_c^2 I_0(x_c),      \\
I^t&=&\eta_2 x_t \lambda_t^2 [ I_0(x_t) +
             y_t \yf I_1(y_t)  +
             x_t \ys (2I_3(x_t,x_H)-8I_4(x_t,x_H)) ], \cont
I^{ct}&=&2 \eta_3 x_c \lambda_t \lambda_c I_6(x_c,x_t),   \nonumber
\eeqa
where $\lambda_q=V_{qs}V_{qd}^*$, and
$\eta_1\sim0.85$, $\eta_2\sim0.62$ and $\eta_3\sim0.36$
are QCD corrections factors~\cite{bbhls}.
Due to the lightness of the valance quarks,
the terms that depend on their masses are neglected.
This implies that the term that may have introduced
\cpv\ from the Higgs
sector, namely, the term that depends on \xy, is very small and cannot
contribute significantly to $\epsilon$.
Our result agrees with~\cite{gil2}.
For the numerical values we use~\cite{pdg,ycpv}
\beq
|\epsilon|=2.26 \times 10^{-3}, \tab
f_K=161~\Mev,   \tab
B_K=2/3 \pm 1/3.
\eeq

The long-distance contributions to $\epsilon$
are negligible in the SM but in MHDM,
where they depend on the phases of the Higgs sector,
they may be important (see~\cite{ycpv,cheng,pok2}).
The largest contribution comes from the
intermediate $\eta^{\prime}$~\cite{dupont,ycpv}
\beq   \label{eq:epsilonld}
|\epsilon_{LD}|\approx 0.02~{\Gev}^2~ \frac{\imxy}{\mhs}
\lef[\ln\frac{\mhs}{m_c^2}-1.5\rit].
\eeq

There is an additional CP violating parameter,
$\epsilon^\prime/\epsilon$.
There is no unambiguous evidence for
$\epsilon^\prime/\epsilon \neq 0$~\cite{pdg}. Furthermore,
there are large uncertainties in its calculation. Therefore,
at present, it is not possible
to extract useful constraints from it.
A discussion can be found in~\cite{barger,buras}.

We conclude that large contribution to $\epsilon$
may arise for large $Y$ through the short-distance contributions,
or for large \imxy\ through the long-distance contributions.

\secb{\dd\ mixing}   \label{se:ddmixing}
The mixing in the $D$ system is different from other neutral
meson systems:
\ben
\item{The intermediate quarks in the box diagrams are down-type.
Consequently, the box diagrams contribution to the mixing is smaller
than in mesons where the intermediate quarks are
up-type~\cite{ddbigi}.}
\item{\dd\ mixing has not been observed, and only
an upper bound has been established~\cite{pdg},
\beq
\Delta m_D<1.5 \times 10^{-13} ~\Gev \tab \nfcl .
\eeq
}
\item{The long distance contribution
to the mixing is probably
dominant over the SM box diagram.
It is estimated to be
about two orders of magnitude below the
experimental bound~\cite{woldon} or even smaller~\cite{georgi}.}
\een
Because of the above reasons, \dd\ mixing constrains
neither the SM parameters \cite{yckm}, nor (as we find)
the MHDM parameters.

The diagrams that contribute are similar to those of
\bb\ mixing (fig.~\ref{fi:bbbox}) with the replacements
$b \rightarrow c$, $d \rightarrow u$ and $t \rightarrow s,b$.
In the calculation
we neglect the SM contribution and the mixed diagrams, while from the
two-Higgs box diagram we take only
the term proportional to $\xf$
\beq
M_{12}=\frac{{G_F}^2}{48{\pi}^2} \mws f_D^2 m_D \ihh{},
\eeq
with
\beq
\ihh{} = \xf x_H [ \lambda_s^2 y_s^2 I_1(y_s)  +
               \lambda_b^2 y_b^2 I_1(y_b)  +
               2 \lambda_b\lambda_s y_b y_s I_5(y_b,y_s) ],
\eeq
where $\lambda_q=V_{cq}V_{uq}^*$.
We can simplify the expression using
\beq
I_1(y_s) \cong I_1(y_b) \cong I_5(y_b,y_s) \cong 1,
\eeq
and get
\beq
M_{12}=\frac{{G_F}^2}{48{\pi}^2} \mws f_D^2 m_D
\xf x_H [ \lambda_s^2 y_s^2  + \lambda_b^2 y_b^2  +
          2 \lambda_b\lambda_s y_b y_s ].  \label{eq:ddmixing}
\eeq
Setting $\lambda_b$ to zero, our result agrees with~\cite{wise},
while setting $m_s$ to zero, our result agrees with~\cite{barger}.
For the numerical values we use \cite{pdg,mati}
\beq
\Delta m_D<1.5 \times 10^{-13} ~\Gev, \tab
f_D=170 \pm 30 ~\Mev.
\eeq

We conclude that large contribution to \dd\ mixing
may arise for large $X$.

\secb{\zbb}
The decay \zbb\ has been measured at LEP.
Within MHDM, extra contributions
arise via virtual scalars in the vacuum polarization diagrams
and the vertex corrections.
Recently, a new method of analyzing the
electroweak precision data was
proposed~\cite{altar}. In this method the non-SM effects
manifest themselves via four observables:
$\epsilon_1$, $\epsilon_2$, $\epsilon_3$ and \epb.
We are interested in \epb\ which is closely related to the
vertex correction to \zbb.
By comparing the experimentally measured quantity, \epb, to the
theoretical calculation of the vertex correction, \db, one can put bounds
on the MHDM parameters~\cite{park}
\beq
\epb \cong 2.3 \db.
\eeq
\db\ is defined in~\cite{finnell}:
\beq
\db = \dbsm + \dbh, \label{eq:vzbb}
\eeq
with
\beqa
\dbsm&\cong&-\frac{20 \alpha}{13 \pi} \lef( \frac{\mts}{m_Z^2} \rit), \\
\dbh&\cong&C \cdot F_L(m_t) \ys \mts, \nonumber
\eeqa
where $C$ is given in \appenc\
and $F_L$ can be found in~\cite{finnell}.
For the numerical data we use~\cite{altar}
\beq
\epb > -1 \times 10^{-2} ~~~\nfcl.
\eeq

There are also terms in \dbh\ proportional to $|m_b X|^2$.
Some of these terms
depend on the neutral Higgs parameters,
but appear with a sign opposite to the terms that depend on the
charged Higgs parameters~\cite{dghk}.
Thus, cancellation may occur
and a bound on $X$ cannot be obtained. This also applies to the
decay \ztautau. There, the MHDM contribution is proportional to
$|m_{\tau} Z|^2$, but the terms that depend on the neutral Higgs
parameters may cancel the terms that depend on the charged Higgs
parameters, and a bound on $Z$ cannot be obtained.

We conclude that large contribution to \zbb\
may arise for large $Y$ or large $X$.
However, a bound can be obtained only on $Y$.

\secb{\bsg}
Recently, the CLEO collaboration has given an upper bound on
the radiative $B$ decay~\cite{cleo}
\beq  \label{eq:nbsg}
{\rm BR}(\bsg)<5.4 \times 10^{-4} ~~~\nfcl,
\eeq
which is only about a factor of 2 above the SM prediction.
Uncertainties in the calculation come mainly from
the value of \mt~\cite{misiak}.
The diagrams that contribute to \bsg\
are given in fig~\ref{fi:fbsg}.
\bef
\ind{dig2}
\caption[fifbsg]
{The additional MHDM diagrams that contribute to \bsg. The SM diagrams
obtained by replacing the \cs\ propagators with a $W$-boson ones.}
\label{fi:fbsg}
\eef
The SM calculation was done in~\cite{lim} and the MHDM calculation
in~\cite{hou,rizzo,pok2,gsw}:
\beq
{\rm BR}(\bsg)=C|\eta_2+G_W(x_t)+(|Y|^2/3)G_W(y_t)
+(XY^*)G_H(y_t)|^2,  \label{eq:bsg}
\eeq
where
\beq
C\equiv \frac{3\alpha\eta_1^2\ {\rm BR}(B\rightarrow X_c\ell\nu)}
{2\pi F_{ps}(m_c^2/m_b^2)}\approx 3\times10^{-4}.
\eeq
$F_{ps}\sim0.5$ is a phase space factor given in \appenc,
$\eta_1\sim0.66$ and
$\eta_2\sim0.57$ are QCD correction factors~\cite{gsw}.
The expressions for the $G$-functions are given in \appenc.
The SM result can be obtained by setting $X$ and $Y$ to zero in
(\ref{eq:bsg}).
In the full calculation~\cite{ma}
a term proportional to $\xs$ appears,
but it is suppressed by $(m_s/m_b)^2$ and we can safely drop it.

We conclude that large contribution may arise for large $Y$ or large \xy.

\secb{\kpinunu}
The \BR\ of the decay \kpinunu\ has not yet been measured.
The current bound is~\cite{e787}
\beq
{\rm BR}(\kpinunu)<6.8 \times 10^{-9} ~~~\nfcl,
\eeq
is about two orders
of magnitude above the SM prediction. The calculation
suffers from uncertainties in \mt\ and $V_{td}$
(for a review see~\cite{bigi}).
The diagrams that contribute to this process are given in
fig.~\ref{fi:fkpinunu}.
\bef
\ind{dig3}
\caption[fifkpinunu]
{The diagrams that contribute to \kpinunu. (a) The $Z$ penguin
diagrams. The solid square
represent the effective $sdZ$ vertex which induced
by diagrams similar to these of \bsg\ (fig.~\ref{fi:fbsg}).
(b) The SM box diagram. (c) The two \css\ box diagram. (d) The
mixed diagram. The second mixed diagram is
not shown.}
\label{fi:fkpinunu}
\eef
In the SM, the penguin and the box diagrams are of the same order, and
intermediate charm and top quarks are important.
In diagrams that involve \css\
the intermediate top quark is dominant.
Since the external quark masses are small,
we consider only terms proportional to internal quark
masses.

The SM diagrams were calculated in~\cite{lim}.
The MHDM result is obtained from the 2HDM
calculation~\cite{barger,buras,geng}
by multiplying the two \css\ box diagram
by $|YZ|^2$ and the mixed diagrams by ${\rm Re}(\yz)$
\beq   \label{eq:brkpinunu}
{\rm BR(\kpinunu)}=\frac{C}{|V_{us}|^2}
\sum_{\ell=e,\mu,\tau} |I^{SM} + I^{HH} + I^{WH} + I^{ZH}|^2,
\eeq
with
\beq
C\equiv\frac{G_F^2 m_W^4}{4\pi^4}
{{\rm BR}(K^+ \rightarrow \pi^0 e^+ \nu)} \approx 6.9 \times 10^{-7}.
\eeq
The SM contribution is $I^{SM}$, the two \css\ box diagram is $I^{HH}$,
the mixed diagrams are $I^{WH}$ and the \chpd\ are $I^{ZH}$
\beqa  \label{eq:ikpinunu}
I^{SM}&=&\eta_c \lambda_c D_{SM}(x_c,x_\ell)+
\lambda_t D_{SM}(x_t,x_\ell), \\
I^{HH}&=&|YZ|^2\lambda_t x_ty_\ell I_5(y_t,y_\ell),  \cont
I^{WH}&=&{\rm Re}(\yz) \lambda_t x_tx_\ell
 (I_7(x_t,x_\ell,x_H)+I_8(x_t,x_\ell,x_H)~),    \cont
I^{ZH}&=&\ys \lambda_t D_{ZH}(x_t,y_t),            \nonumber
\eeqa
where $\lambda_q=V_{qd}V_{qs}^*$
and the $D$-functions are given in \appenc.
$\eta_c \sim 0.7$ is a QCD correction factor~\cite{bigi},
while QCD corrections
to the diagrams with an intermediate top quark are small~\cite{buch1}.
While for the penguin diagrams the MHDM contribution is independent
of the final lepton, the \chbd\ depend on the internal lepton mass and
it may be important only for internal $\tau$.
In 2HDM (\modb) $\yz=1$ and the \chbd\ are negligible.
Our results disagree with those of~\cite{bigi} where it is
claimed that the \chbd\ are not important even
within MHDM. We believe that the reason for this is
that they assume that all the coefficients in the Higgs potential are
of the same order. Then, $|YZ|=O(1)$ and the
effect is indeed small.

A similar calculation applies to the decay \bknunu.
However, due to the lack of experimental
data, no constraints on the parameters can be obtained.

We conclude that large contribution to \kpinunu\
may arise for large $Y$ or large \yz.

\secb{\bxtaunu}
The \BR\ for \bxtaunu\ has been recently measured by the ALEPH
collaboration~\cite{aleph} which leads to the upper bound
\beq \label{eq:nbtauex}
{\rm BR}(\bxtaunu) <4.00 \% ~~~\nfcl.
\eeq
The uncertainties in the calculation are mainly from $m_b$ and $m_c$.
The diagrams that contribute to the decay are given in
fig.~\ref{fi:fbxtaunu}.
\bef
\ind{dig4}
\caption[fifbxtaunu]
{The diagrams that contribute to \bxtaunu.
(a) The SM diagram. (b) The \cs\ mediated diagram.}
\label{fi:fbxtaunu}
\eef
In the calculation we neglect charmless final states, since they are
highly suppressed by CKM elements.
A recent SM calculation was done in~\cite{flnn}.
The MHDM result is obtained from
the 2HDM calculation~\cite{pokor,kalino,isidori}
by replacing $\xs$ with \xz
\beq
BR(\bxtaunu)=BR^{SM}(\bxtaunu)(1+\frac{1}{4}|R|^2-D \cdot {\rm Re}(R)),
\label {eq:bxtaunu}
\eeq
where the SM result is~\cite{flnn}
\beq   \label{eq:nbtausm}
BR^{SM}(\bxtaunu)=
(2.30 \pm 0.25) \%,
\eeq
and
\beq
R\equiv \frac{\mtau \mb \xz}{\mhs}, \tab
D\equiv
2\frac{\mtau}{m_b}
\frac{\tilde{F}_{ps}^{int}(m_c/\mb,m_{\tau}/\mb)}
     {\tilde{F}_{ps}(m_c/\mb,m_{\tau}/\mb)}.
\eeq
The phase space functions
$\tilde{F}_{ps}$ and $\tilde{F}_{ps}^{int}$ are given in
\appenc. For the ranges of masses that we use
$D \approx 0.43 \pm 0.01$.
We disagree with~\cite{isidori} about the numerical value of $D$.
Due to this disagreement the bound that we obtain is weaker than
the bound obtained by~\cite{isidori}.
In 2HDM (\modb) the interference
term always reduces the rate, while in MHDM it may also enhance it,
depending on the relative phase between $X$ and $Z$.
Nevertheless, in order to get bounds on $|XZ|$ one should take the
minimal value for the \BR,
which corresponds to $\arg(\xz)=0$. Thus, the bound on
$\xs$ in 2HDM is the same as the bound on $|XZ|$ in MHDM.

We conclude that large contribution to \bxtaunu\
may arise for large $XZ$.

\secb{$e$--$\mu$ universality in $\tau$ decays}   \label{se:lepuni}
$e$--$\mu$ universality is confirmed in leptonic decays of pions and
kaons, and in $\tau$ decays (for a review
see~\cite{wstro}). The SM and MHDM diagrams are similar to those of
\bxtaunu\ (fig.~\ref{fi:fbxtaunu}) with the replacements:
$b \rarrow \tau$,
$c \rarrow \nu_{\tau}$ and
$\tau \rarrow \ell$ ($\ell=e,\mu$).
At first
glance, it seems that a useful bound on the MHDM
parameters cannot be obtained since the final leptons are very light.
However, in purely leptonic decays
both the theoretical calculation and the experimental measurement
are clean. We then require that the \cs\ contribution does not
exceed the difference between the SM prediction and the experimental
result.
The calculation in the case of muon decay was done in~\cite{haber}
and for $\tau$ decay
in~\cite{pokor,holik}\footnote{We disagree with the
numerical factor of the interference term in eq.~(8) of~\cite{holik}.
It seems to us that they divide by $F_{ps}^{int}/F_{ps}$ instead of
multiplying by this factor.}
\beq
\frac{{\rm BR}(\taumu)}{{\rm BR}(\taue)}=F_{ps}(m_{\mu}^2/m_{\tau}^2)
[1+\frac{R^2}{4}-D \cdot R]
\cong 0.9726[1+\frac{R^2}{4}-0.125R],  \label{eq:emuuni}
\eeq
where
\beq
R\equiv\frac{\mtau \mmu \zs}{\mhs}, \tab
D\equiv
2\frac{\mmu}{\mtau}\frac{F_{ps}^{int}(m_{\mu}^2/m_{\tau}^2)}
{F_{ps}(m_{\mu}^2/m_{\tau}^2)}.
\eeq
$F_{ps}^{int}$,
the phase space function for the interference term, is
given in \appenc.
We use the recent numerical values~\cite{roney},
which leads to the  upper bound
\beq \label{eq:nemu}
\frac{{\rm BR}(\taumu)}{{\rm BR}(\taue)F_{ps}(m_{\mu}^2/m_{\tau}^2)}
<1.035 ~~~\nfcl.
\eeq

We conclude that strong deviation from
$e$--$\mu$ universality in $\tau$ decays
may arise for large $Z$.

\secb{Electric dipole moment of the neutron}
An electric dipole moment (EDM) of an elementary particle is a
manifestation of CP violation (for a review see~\cite{ycpv}). In our
discussion we concentrate on the EDM of the neutron (NEDM), $D_n$.
Although there are very large uncertainties in the calculation, we know
that the contribution to the NEDM from the electro-weak sector of the
SM is much smaller than the current experimental
bound~\cite{pdg}
\beq
|D_n| \le 1.2 \times 10^{-25}~~ e~{\rm cm}~~~\nfcl.
\eeq
A large NEDM can be generated within MHDM in many ways (for a review
see~\cite{cheng}). In our discussion we concentrate on the
two that involve \css: the EDM of
the down quark, and the three gluon operator.
The contribution to the NEDM from the EDM of the down
quark is given in~\cite{beall}. We neglect the
contribution from the internal top quark and using $m_c \ll m_H$
we get
\beq
D_n=\frac{\sqrt{2}G_F}{9\pi^2} m_d~\imxy |V_{cd}|^2
\lef(\ln(y_c)+0.75 \rit)y_c.  \label{eq:nedm1}
\eeq
For the three gluon operator we use the naive
dimensional analysis result~\cite{dicus}
\beq
D_n=\frac{4\sqrt{2} \zeta G_F g_s^3}{(4\pi)^{4}}\imxy
h^{\prime}(y_t),  \label{eq:nedm2}
\eeq
where
\beq
h^{\prime}(y)=\frac{y}{8{(1-y)}^3} \lef[ 4y-y^2-3-2\ln(y) \rit].
\eeq
$\zeta$, the QCD correction factor, depends sensitively on the
scale~\cite{ckly}
\beq
\zeta=10^{-3} - 10^{-1}.
\eeq

We conclude that large contribution to the NEDM
may arise for large \imxy.

\secb{\bll}
The \BR\ for \bll\ has not been measured, but there
is an upper bound for $\ell = e,\mu$.
The diagrams that contribute to \bll\
are similar to the diagrams that contribute to \kpinunu\
(fig.~\ref{fi:fkpinunu}).
The photon penguin diagrams vanish because of
electromagnetic current conservation. The \chpd\
contribute universally for all leptons. However, the \chbd\ depend on
the lepton mass. Thus, it is important only for \btautau.
We emphasize that in 2HDM the box diagram is
negligible since it is proportional to
$|YZ|^2$, which is 1 in \modb, or $O(1)$ in \moda.
A recent
SM calculation was done in~\cite{buch2} and the 2HDM in~\cite{hnr}.
We calculated the box diagram. We neglect terms
proportional to external quarks masses, and the QCD
correction which is small~\cite{buch2}.
We get
\beq
{\rm BR}(\bll)={\rm BR}_{SM}(\bll)
\lef[1+\frac{G^H_Z(x_t,y_t,y_{\ell})-G^H_{box}(x_t,y_t,y_{\ell})}
{G^{SM}(x_t)}\rit]^2,
\eeq
where
\beq
{\rm BR}_{SM}(\bll)={\tau}_B\frac{G_F^4 m_W^4}{32 {\pi}^5} f_B^2 m_B
m_{\ell}^2 \sqrt{1-\frac{4m_{\ell}^2}{m_B^2}}
\lef| G^{SM}(x_t) {\lambda}_t \rit|^2,
\eeq
$\lambda_q=V_{qd}V_{qb}^*$ and
$G^H_Z$ and $G^H_{box}$ are the charged scalar penguin and box
diagram contributions, respectively:
\beqa
G^H_Z&=&y_t x_t \ys J(y_t),  \\
G^H_{box}&=&\frac{1}{8}x_t y_\ell (|YZ|^2  I_9(y_t) +
2{\rm Re}(YZ^*) I_{10}(x_t,x_H) ).   \nonumber
\eeqa
$G_{SM}$, $J$ and $I_i$ are given in \appenc.
Our results agree with~\cite{he}, but
disagree with~\cite{hnr} about the Dirac structure of
the matrix element, and
the sign of the first term of $I_9$. However, since they work within
the framework of
2HDM, where the box diagram is negligible, this disagreement does
not have numerical significance.

A few points are in order:
\begin{enumerate}
\item{The same calculation applies to $B_s$ decays with
($d \rightarrow s$). Consequently, the \BR\ for $B_s$ decay
is larger by approximately $|\frac{V_{ts}}{V_{td}}|^2$.}
\item{It was noted~\cite{he,skiba}
that large effects may arise
due to neutral scalar penguin diagrams, even in 2HDM. This
contribution depends
on neutral Higgs sector parameters and could also be small.
We note that the
contributions from the \cs\ and from the neutral scalars are
independent.}
\item{The process \kmumu\ gets contributions from
diagrams similar to \bll. Since the muon is light, the
box diagram is negligible.
In addition, the long distance contribution dominates.
Thus, it is difficult to estimate the \chpd\ contribution.
We do not discuss this process.
For a detailed discussion in the framework of
2HDM, see~\cite{barger}.}
\item{The processes \bxll\ gets
contributions from diagrams similar to \bll,
with the addition of photon penguin diagrams.
In 2HDM the \chbd\ are negligible~\cite{hou,desh}.
However, in MHDM they may be important for final $\tau$ leptons.
Then an interesting situation arises:
The Z penguin depends on $Y$,
the photon penguin depends on $Y$ and \yx,
and the \chbd\ depend on $|YZ|$. The process \bxtautau\ then
depends on {\em all} the charged Higgs
parameters. Furthermore, the neutral scalar
penguin~\cite{he,skiba} may also contribute.
However, due to the poor experimental data, no bounds can be
obtained from these decays.}
\end{enumerate}

We conclude that large contribution to \bll\
may arise for large $Y$, and for \btautau\ also for large $YZ$.

\secb{\blnu}
The branching ratios of the decays \blnu\ ($\ell=\mu$,$\tau$)
have not yet been measured, and the
current \ncl\ bounds~\cite{cleoblnu}
\beqa
{\rm BR}(\bmunu)&<&2.0 \times 10^{-5} , \\
{\rm BR}(\btaunu)&<&1.2 \% , \nonumber
\eeqa
are about two orders
of magnitude above the SM predictions.
The calculation suffers from uncertainties in $f_B$ and $V_{ub}$.
The diagrams that contribute to the decays are similar to those of
\bxtaunu\ (fig.~\ref{fi:fbxtaunu}),
with the replacement $c \rightarrow u$.
The MHDM result is obtained from the
2HDM calculation~\cite{hou2}
by replacing $\xs$ by \xz:
\beq
BR(\blnu)=BR^{SM}(\blnu)|1-R|^2, \label{eq:blnu}
\eeq
where the SM result is
\beq
BR^{SM}(\blnu)=
\frac{G_F^2 m_B m_{\ell}^2}{8 \pi} \lef( 1-\frac{m_{\ell}^2}{m_B^2} \rit)
f_B^2 |V_{ub}|^2 \tau_B,
\eeq
and
\beq
R \equiv \frac{m_B^2 \xz}{\mhs}.
\eeq
In 2HDM (\modb) the interference
term always reduces the rate, while in MHDM it may also enhance it,
depending on the relative phase between $X$ and $Z$.
Nevertheless, in order to get bounds on $|XZ|$ one should take the
minimal value for the \BR,
which corresponds to $\arg(\xz)=0$. Thus, the bound on
$\xs$ in 2HDM is the same as the bound on $|XZ|$ in MHDM.

We conclude that large contribution to \blnu\
may arise for large $XZ$.

\seca{Combined analysis}
In this chapter we combine the results from the previous chapter
and determine the region in parameter space
allowed by the data.

Several processes are sensitive to $|Y|$.
For some of them there are very poor experimental data:
\bsbs\ mixing and \bknunu.
For others, the experimental bounds are far above the
SM predictions: \kpinunu, \bll\ and \bxll.
The bounds obtained from $\Delta m_K$ and \kmumu\ are very weak.
The decay \bsg\ gives a strong constraint
on $|Y|$ in the case of 2HDM
(\modb)~\cite{hewett}. In MHDM, however, there is
a possibility of cancellation between the two terms that depend on the
MHDM parameters (see eq.~(\ref{eq:bsg})),
and such a bound cannot be obtained.
Then, there are three observables that constrain $|Y|$:
$\epsilon$ (eq.~(\ref{eq:epsilon})),
$x_d$ (eq.~(\ref{eq:monetwo})) and \epb\ (eq.~(\ref{eq:vzbb})).
At present, \epb\ gives the strongest bound~\cite{park}.
It is important to note that the bound from \epb\ is strong since
the experimental value is far from the SM prediction. Actually, at
the $1 \sigma$ level \mt\ is found to be less than the CDF lower bound.
Had the data fitted large \mt, the bound on $|Y|$ from \bb\
mixing may become the strongest.
Our results are given in fig.~\ref{fi:cy}.
\begin{figure}[t]
\centerline{\zpsfig{file=zbb.eps}}
\caption[ficy]{The upper bound on $|Y|$ as a function of the lightest
\cs\ mass $m_H$.
The three curves correspond to
$m_t=100$ (solid), 140 (dashed) and 180 (dotted) GeV.}
\label{fi:cy}
\end{figure}
For example, we obtain
\beq
|Y|<1.3 \tab \mbox{for \mt=140 GeV and $\mh=45$ GeV.}
\eeq

The only potential constraint on $|X|$ comes from \dd\ mixing
(eq.~(\ref{eq:ddmixing})).
However, the current experimental bound with
the large uncertainties in the CKM elements gives no useful bound.
We conclude that $|X|$ is constrained by the
perturbativity bound (eq.~(\ref{eq:pert})):
\beq
|X|<130.
\eeq

$|Z|$ is constrained by
$e$--$\mu$ universality in $\tau$ decays (eq.~(\ref{eq:emuuni}))
and by the perturbativity bound (eq.~(\ref{eq:pert})).
For $m_H > 175$ GeV the perturbativity bound is stronger:
\beq
|Z|<\min(1.93 ~m_H {~\Gev}^{-1},340).
\eeq

$|XZ|$ is constrained by \bxtaunu\ (eq.~(\ref{eq:bxtaunu})),
\bmunu, \btaunu\ (eq.~(\ref{eq:blnu}))
and by the product
of the bounds on $|X|$ and $|Z|$.
For light \cs\ \bxtaunu\ give the strongest bound, while for
$m_H > 370$ GeV the perturbativity bound is the strongest:
\beq  \label{eq:boundxz}
|XZ|<\min(0.32 ~\mhs {~\Gev}^{-2},44200).
\eeq
In 2HDM (\modb), where $X=Z$,
eq.~(\ref{eq:boundxz}) gives
$|X|<0.56 ~m_H {~\Gev}^{-1}$, which is weaker than the published
bound~\cite{aleph,isidori}. The difference arises due to
the different values calculated for the interference term and
the different values taken for the SM \BR.

It is important to note that
below certain values of $|XZ|$ and \mh\ the destructive
interference leads to suppression and a bound cannot be
obtained~\cite{hou2}.
This happens at
$|XZ| < 0.18 ~\mhs {~\Gev}^{-2}$ for \bxtaunu\ and at
$|XZ| < 0.07 ~\mhs {~\Gev}^{-2}$ for \blnu.
We see that although the current bound from \bxtaunu\ is stronger,
\blnu\ can potentially give a stronger bound.

The bound on Re($XZ^*$) is the same as the bound on $|XZ|$.
However, the bound on the CP violating parameter \imxz\ is stronger.
\imxz\ can be bounded from the same processes that bound $|XZ|$.
For light \cs\ \bxtaunu\ give the strongest bound, while for
$m_H > 440$ GeV the perturbativity bound is the strongest:
\beq  \label{eq:boundimxz}
\imxz<\min(0.23 ~\mhs {~\Gev}^{-2},44200).
\eeq

Several processes are sensitive to $|YZ|$.
For some of them there are very poor experimental data:
\bknunu, \btautau\ and \bxtautau. The
only measurement that could potentially bound $|YZ|$ is
\kpinunu\ (eq.~(\ref{eq:brkpinunu})).
However, due to
the current experimental bound and the large uncertainties in the
CKM elements this bound is weaker than
the product of the bounds on $|Y|$ and $|Z|$.
The bounds on Im($YZ^*$) and on Re($XY^*$)
are the same as the bound on $|YZ|$.
For example, we obtain
\beq
|YZ|<110 \tab \mbox{for \mt=140 GeV and $\mh=45$ GeV.}
\eeq

The processes that are sensitive to \xy\ are \bb\ mixing,
\bxll\ and \bsg.
The strongest bound comes from \bsg\ (eq.~(\ref{eq:bsg})).
The bound on $|XY|$ is very sensitive to $\arg(\xy)$ and $|Y|$.
The weakest upper bound on $|XY|$
is obtained for maximal $|Y|$ and
$\arg(\xy)=\pi$. Then, the cancellation between the two terms,
the one proportional to \xy\ and the one
proportional to $\ys$, is maximal.
Our results are given in fig.~\ref{fi:cxy}.
\bef[t]
\centerline{\zpsfig{file=ft2.eps}}
\caption[ficxy]{
The upper bound on $|XY|$ as a function of the lightest
\cs\ mass $m_H$.
The three curves correspond to
$m_t=100$ (solid), 140 (dashed) and 180 (dotted) GeV.}
\label{fi:cxy}
\eef
For example, we obtain
\beq
|XY|<4 \tab \mbox{for \mt=140 GeV and $\mh=45$ GeV.}
\eeq

The bound on Re($XY^*$) is the same as that on $|XY|$.
However, the bound on the CP violating parameter \imxy\ is stronger.
\imxy\ is bounded by CP violating processes:
NEDM  (eq.~(\ref{eq:nedm1}))
and $\epsilon$ (eq.~\ref{eq:epsilonld})),
and by CP conserving ones: \bb\ mixing, \bxll\ and \bsg.
The strongest bound comes from \bsg~\cite{yuval}.
Our results are given in fig.~\ref{fi:cimxy}.
\bef[t]
\centerline{\zpsfig{file=ft3.eps}}
\caption[ficimxy]{
The upper bound on \imxy\ as a function of the lightest
\cs\ mass $m_H$.
The three curves correspond to
$m_t=100$ (solid), 140 (dashed) and 180 (dotted) GeV.}
\label{fi:cimxy}
\eef
For example, we obtain
\beq
\imxy<2 \tab \mbox{for \mt=140 GeV and $\mh=45$ GeV.}
\eeq

\seca{Discussion}
In this chapter we discuss the differences between the
general MHDM and the widely discussed 2HDM (\modb)\footnote{In
this chapter we refer only to \modb\ of 2HDM.}. In 2HDM
there are only two parameters in the charged Higgs sector: the
mass of the \cs, \mh, and the ratio of the vacuum expectation values,
\tgb. The lower bound
on \tgb\ is the same as the bound on $1/|Y|$ obtained from the
decay \zbb. Since in 2HDM $X=Z$, the  upper bound on \tgb\
is the same as the bound on $\sqrt{|XZ|}$. For intermediate values
of \tgb\ the bound on \mh\ is obtained from the decay \bsg\
(see \eg~\cite{misiak}).
There are two main reasons for the differences
between MHDM and 2HDM:
\bei
\item{In MHDM there are four parameters in the charged Higgs sector.
In 2HDM, $\xy=1$ and $X=Z$ so there are only two.
Thus, processes that involve combinations of these parameters can
be enhanced or suppressed in MHDM compared to 2HDM.}
\item{The bounds on the parameters in MHDM may be weaker than
the corresponding bounds in 2HDM.
Thus, the possible effects are larger.}
\eei

\secb{Bound on \mh}
In 2HDM \bsg\ gives a lower bound on \mh\
almost independent of \tgb:
$\mh \raise.3ex\hbox{$>$\kern-.75em\lower1ex\hbox{$\sim$}} 100$ GeV.
In MHDM,
however, this bound does not hold. This bound has implications for
the experimental search of the \cs. In LEP200, a \cs\ of mass up to
approximately 80 GeV can be found~\cite{ron}.
Consequently, the \cs\ of
2HDM cannot be detected at LEP200 while the \cs\ of MHDM can.
Furthermore, the \cs\ of 2HDM decays dominantly to \tnu\ and \scb.
Within MHDM, when $|X|$ is large and $|Y|$ and $|Z|$ are small,
the \bc\ channel becomes important. Large background from
$W$ decays is expected if $\mh \sim m_W$. Since the $W$-boson
hardly decays
into \bc, the $H^+ \rightarrow c \bar{b}$
decay mode is easier to detect.

\secb{Processes that depend on $|X|$ or $|Z|$}
In 2HDM the bound on $X=Z$ is the same as the bound on
$\sqrt{|XZ|}$ in MHDM. Consequently, the bounds on $|X|$ and
$|Z|$ are weaker within MHDM. Thus, processes that depend on
$|X|$ or $|Z|$ may be enhanced within MHDM compared to the
2HDM case:
\bei
\item{
The \chbd\ can saturate the experimental bound on \dd\ mixing.
Furthermore, measurable CP violating effects
in the interference of mixing and decay may arise.
Intuitively, the reason is that
the \chbd\ introduce dependence on third generation parameters.
Based on the discussion in~\cite{ddbigi} we find that
the measured asymmetry
\beq
A_{CP} \equiv
\frac{N[(\ell^-X)_D(P^+P^-)_D]-N[(\ell^+X)_D(P^+P^-)_D]}
{N[(\ell^-X)_D(P^+P^-)_D]+N[(\ell^+X)_D(P^+P^-)_D]}
\eeq
($P^+P^-$ is a CP eigenstate, \eg\ $\pi^+ \pi^-$ or $K^+K^-$)
can reach its current upper bound
\beq
A_{CP} \leq 0.16.
\eeq
Note, that no new source of \cpv\ beyond the
phase of the CKM matrix is required.
A recent estimate~\cite{bigipri} shows that within
one year operation of a Tau-Charm Factory
asymmetries of order 1\% are measurable.
}
\item{
\bsbs\ mixing can be enhanced.
The ratio between the two mixing
observables, $x_d$ and $x_s$, is given by:
\beq \label{eq:xsxd}
\frac{x_s}{x_d} = r \cdot \frac{{|V_{ts}|}^2}{{\vtd}^2}.
\eeq
Deviations from $r=1$ in eq.~(\ref{eq:xsxd}) are due
to flavor $SU(3)$ breaking effects.
{}From eq.~(\ref{eq:iwwhh}) we see that when $|X|$ is close to its upper
bound and $m_H$ to its lower bound, the last term in $I_{HH}$
dominates $x_s$ but it is still small for $x_d$.
Then, $r$ is enhanced. We find that even $r \sim 7$ is possible.
}
\item{
Violation of $e$--$\mu$ universality in $\tau$ decays
is proportional to $|Z|^2$. Since the bound
on $|Z|$ is weaker in MHDM compared with 2HDM, the violation
can be stronger.}
\eei

\secb{Processes that depend on \yz\ or \xy}
While in 2HDM
$\xy=\yz=1$, in MHDM \xy\ and \yz\ are, in principle, arbitrary
complex numbers.
Thus, processes that involve these combinations can
be enhanced or suppressed in MHDM compared with 2HDM:
\bei
\item{
FCNC decays with third generation leptons in the final
state.
In such processes the \chbd\ may be important.
They depends on \yz\ and on the lepton mass.
Thus,
in processes where leptons from the first and second generation
are involved the box diagram is negligible.
For \kpinunu\ we find that the SM upper bound
\beq
\brsm (\kpinunu)<4 \times 10^{-10},
\eeq
is modified into
\beq
\brmh (\kpinunu)<1.4 \times 10^{-9},
\eeq
which is more than 3 times weaker than the SM bound.
For \bknunu\ we find that the SM upper bound
\beq
\brsm (\bknunu)<5.5 \times 10^{-5},
\eeq
is modified into
\beq
\brmh (\bknunu)<8.5 \times 10^{-4},
\eeq
which is more than a order of magnitude weaker than the SM bound.
A similar enhancement may occur for the decays
\btautau, \bstautau\ and \bxtautau.}
\item{Processes that involve the photon penguin diagrams
depend on \xy. In 2HDM, the intermediate
\cs\ enhances the
effective $bs \gamma$ vertex. In MHDM, the intermediate \cs\
may cause larger enhancement or,
when $\arg(\xy) \sim \pi$,  suppress the
effective $bs \gamma$ vertex. Thus, processes like \bsg\ and \bxll\
may have different predictions in MHDM and in 2HDM.}
\eei

\secb{\cpv}
Within MHDM additional CP violating
phases in charged scalar exchange are allowed. Those phases
are absent in 2HDM. \cpv\ from charged scalar exchange
can make only a very small contribution to $\epsilon$ (less then 4\%),
and cannot be the {\it only} source of CP violation~\cite{pok2}.
The  charged scalar exchange may have a small effect on
CP asymmetries in neutral $B$ decays, at most 0.02 shift in the
measured CP asymmetries~\cite{yuval}. However, large  contributions
to the NEDM and to \cpv\ in top decays~\cite{eilam} are possible.

\seca{Summary}
We have studied the charged Higgs sector of the general
Multi-Higgs-Doublet Model (MHDM). Using experimental data
and the requirement of perturbativity we constrained
the model parameters. These bounds are summarized in Table~\ref{tb:sum}
for a representative value of $m_t=140$~GeV.
\begin{table}
\begin{center}
\begin{tabular}{||c||c|c|c|c|c|c|c|c||}
\hline
$m_H$ [GeV] & $|X|$ & $|Y|$ & $|Z|$ &
$|XY|$ & $|XZ|$ & $|YZ|$ & \imxy & \imxz \\
\hline
45 & 130 & 1.3 & 85 & 4 & 650 & 110 & 2 & 465 \\
\hline
200 & 130 & 1.9 & 340 & 8 & 12800 & 650 & 4 & 9200 \\
\hline
\end{tabular}
\end{center}
\caption[tbsum]{The upper bounds on the MHDM parameters
for a representative value of $m_t=140$~GeV.}
\label{tb:sum}
\end{table}

We pointed out differences between the
general MHDM to the widely discussed 2HDM (\modb):
\bei
\item{The bound on the \cs\ mass is lower in MHDM. Thus,
the \cs\ of MHDM are in the mass range accessible by LEP200.}
\item{The bounds on $|X|$ and $|Z|$ are weaker in MHDM. Then,
charged scalar
exchange can saturate the experimental bound on \dd\ mixing,
\bsbs\ can be enhanced and strong
violation from lepton universality is possible.}
\item{Large effects are possible in FCNC
decays with third generation leptons.
In these processes the \chbd\ are
important only in MHDM.}
\item{There are differences in the predictions for processes that
involve the photon penguin diagrams, namely, \bsg\ and \bxll.}
\item{In 2HDM there is no source
of \cpv\ from the Higgs sector. In MHDM there are new CP violating
sources but only when the \css\ are not degenerate.
Large \cpv\ effects, that arise from the CKM matrix, are possible
in $D$ decays.}
\eei

We conclude that MHDM may have large effects in various processes.
Many of these effects do not appear in the 2HDM.

\seca*{Acknowledgments}
I thank Yossi Nir for his help throughout this work,
Miriam Leurer and Zoltan Ligeti for useful conversations.
I am grateful to Donald Finnell for
providing me a computer program that enabled the calculation of
\db.

\appendix
\seca{Proofs of relations}
\setcounter{prop}{0}
In this appendix we give proofs of relations between the
coupling constants of MHDM.
\begin{prop}   \label{th:ap1}
\beqa   \label{eq:thap1}
\sum_{i=2}^n X_iY_i^*=1& & \mbox{{\rm(for $\Phi_d\not= \Phi_u$)}}.
\eeqa
\end{prop}
For $\Phi_d \neq \Phi_u$,
unitarity of $U$ gives
\beq
\sum_{i=1}^n U_{di}U^*_{iu}=0 ~~\Rightarrow
{}~~\sum_{i=2}^n U_{di}U^*_{iu}=-U_{d1}U^*_{1u} ~~\Rightarrow
{}~~\frac{\sum_{i=2}^n U_{di}U^*_{iu}}{U_{d1}U^*_{1u}}=-1.
\eeq
Using the definition of $X_i$ and $Y_i$ (eq.~(\ref{eq:defxyz}))
we get eq.~(\ref{eq:thap1}).
The proof of the other two relations in eq.~(\ref{eq:th1}) is obtained
in a similar way.

\begin{prop} \label{th:ap2}
\beqa
\sum_{i=2}^n {|X_i|}^2&=& \frac{v^2}{v_d^2}-1.
\eeqa
\end{prop}
Using eq.~(\ref{eq:defxyz}), eq.~(\ref{eq:defvi}) and the unitarity
of $U$ we get
\beq
\sum_{i=2}^n {|X_i|}^2=
\frac{\sum_{i=2}^n {|U_{di}|}^2}{{|U_{d1}|}^2}=
\frac{\sum_{i=1}^n {|U_{di}|}^2}{{|U_{d1}|}^2}-1=
\frac{1}{{|U_{d1}|}^2}-1=
\frac{v^2}{v_d^2}-1.
\eeq
The proof of the other two relations in eq.~(\ref{eq:th2}) is obtained
in a similar way.

\begin{prop} \label{th:ap3}
{\rm $X_i$, $Y_j$ and $Z_k$ $(i,j,k\in 2..n)$
cannot be simultaneously large
(more than $\sqrt{n-1}$), unless there is at
least one Higgs doublet that does not couple to fermions.}
\end{prop}
The proof is given for a general case where there are $m$ types of
massive fermions and $n$ Higgs doublets.
We denote the coupling of each fermion type by $X^k$, where $k$ is
the fermion type. We demand
\beq
|X_i^k|>a.
\eeq
Using eq.~(\ref{eq:th2})
we get
\beq
a^2 < {|X_i^k|}^2 \leq
\sum_{i=2}^n {|X_i^k|}^2= \frac{v^2}{v_k^2}-1  ~~\Rightarrow
{}~~v^2 > (a^2+1)v_k^2.
\eeq
Summing over all fermions types we get
\beq \label{eq:aat3}
mv^2>(a^2+1)\sum_{k=1}^m v_k^2.
\eeq
We have to distinguish between two cases. First,
$m=n$ and each fermion type couples exactly to one Higgs doublets.
Then eq.~(\ref{eq:aat3}) with eq.~(\ref{eq:defv}) lead to
\beq
nv^2>(a^2+1)v^2 ~~\Rightarrow ~~a<\sqrt{n-1}.
\eeq
On the other hand, if
there is at least one Higgs doublet that does not couple to fermions,
$a$ is not bounded.

\seca{Formulae}
In this appendix we give the expressions of various functions
that play a role in our study.

\secb{Neutral meson mixing}
All the integrals that we have to
calculate are of the form
\beq
J_n(a,b,c,d)=\int_0^{\infty} \frac{x^n~dx} {(x+a)(x+b)(x+c)(x+d)},
\eeq
for $n=0,1,2$. We get
\beqa
J_n&=&\frac{a^n~\ln(a)}{(a-b)(a-c)(a-d)} +
      \frac{b^n~\ln(b)}{(b-a)(b-c)(b-d)} + \\
& &   \frac{c^n~\ln(c)}{(c-a)(c-b)(c-d)} +
      \frac{d^n~\ln(d)}{(d-a)(d-b)(d-c)}. \nonumber
\eeqa
The integrals in the text are
\beqa
I_1(y_t)&=&J_2(1,1,y_t,y_t) =
\frac{1+y_t}{{(1-y_t)}^2}+\frac{2y_t\ln(y_t)}{{(1-y_t)}^3}, \\
I_2(y_t)&=&J_1(1,1,y_t,y_t) =
\frac{2}{{(1-y_t)}^2}+\frac{(1+y_t)\ln(y_t)}{{(1-y_t)}^3}, \cont
I_3(x_t,x_H)&=&J_2(1,x_H,x_t,x_t)= \cont
& &\frac{x_t}{(x_t-x_H)(1-x_t)}+
\frac{x_H^2\ln(x_H)}{(1-x_H){(x_t-x_H)}^2} +
\frac{x_t(x_t+x_tx_H-2x_H)\ln(x_t)}{{(1-x_t)}^2{(x_t-x_H)}^2}, \cont
I_4(x_t,x_H)&=&J_1(1,x_H,x_t,x_t)= \cont
& &\frac{1}{(x_t-x_H)(1-x_t)}+
\frac{x_H\ln(x_H)}{(1-x_H){(x_t-x_H)}^2} +
\frac{(x_t^2-x_H)\ln(x_t)}{{(1-x_t)}^2{(x_t-x_H)}^2}, \cont
I_5(y_b,y_s)&=&J_2(1,1,y_s,y_b). \nonumber
\eeqa
The SM integral can also be
obtained from the general integral $J$. However,
we keep the conventional notation~\cite{ycpv}
\beqa
I_0(y)&=&1-\frac{3y(1+y)}{{4(1-y)}^2}
\lef[1+\frac{2y~\ln(y)}{1-y^2} \rit], \\
I_6(x,y)&=&\ln(y/x)-\frac{3y}{4(1-y)}
\lef[1+\frac{y~\ln(y)}{1-y} \rit]. \nonumber
\eeqa
For the calculations of the matrix elements
we use the Vacuum Insertion Approximation
\beqa
\vll&=&\viavll=\frac{1}{3} f_B^2 m_B , \\
\sll&=&\viasll=-\frac{5}{6} \left(\frac{m_B}{m_b+m_d}\right)^2
f_B^2 m_B. \nonumber
\eeqa

\secb{\zbb}
\beq
C=\frac{\alpha~v_L}{2 \pi \sws (v_L^2 + v_R^2)},
\eeq
with
\beq
v_L=-\frac{1}{2}+\frac{1}{3}\sws, \tab
v_R=\frac{1}{3}\sws.
\eeq

\secb{\bsg}
\beqa
G_W(x)&=&\frac{x}{12{(1-x)}^4} \lef[
(7-5x-8x^2)(1-x)+6x(2-3x)\ln(x) \rit], \\
G_H(x)&=&\frac{-x}{6{(1-x)}^3} \lef[
(3-5x)(1-x)+2(2-3x)\ln(x) \rit], \cont
F_{ps}(x)&=&1-8x+8x^3-x^4-12x^2\ln(x). \nonumber
\eeqa

\secb{\kpinunu}
\beqa
D_{SM}(x,y)&=&\frac{1}{8} \lef \{
\frac{xy}{y-x}{\lef(\frac{4-y}{1-y}\rit)}^2~\ln(y) -2x
-\lef(1-\frac{3}{1-y}\rit)\frac{3x}{1-x} \rit. \\
& & \lef. -\lef[\frac{x}{y-x}{\lef(\frac{4-x}{1-x}\rit)}^2~
  +1 +\frac{3}{{(1-x)}^2}\rit] x~\ln(x) \rit\}, \cont
D_{ZH}(x,y)&=&\frac{xy}{4}~\frac{1-y+\ln(y)}{{(1-y)}^2}, \cont
I_7(x_t,x_\ell,x_H)&=&J_2(1,x_t,x_\ell,x_H), \cont
I_8(x_t,x_\ell,x_H)&=&J_1(1,x_t,x_\ell,x_H). \nonumber
\eeqa

\secb{\bxtaunu}
The phase space factors for three body decay with only one massless
anti-particle in the final state are~\cite{cortes,kalino}
\beqa
\tilde{F}_{ps}(x,y)&=&12\int_{{(x+y)}^2}^1 \frac{ds}{s}
(s-x^2-y^2) {(1-s)}^2 g(s,x^2,y^2), \\
\tilde{F}_{ps}^{int}(x,y)&=&6\int_{x^2}^{{(1-y)}^2} \frac{ds}{s}
{(s-x^2)}^2 g(1,s,y^2), \nonumber
\eeqa
with
\beq
g(a,b,c)=\sqrt{a^2+b^2+c^2-2(ab+ac+bc)}.
\eeq
We do not give the analytic expressions of these functions.
$\tilde{F}_{ps}$ can be found in~\cite{cortes}, while
we use a computer program
to calculate $\tilde{F}_{ps}^{int}$.

\secb{$e$--$\mu$ universality in $\tau$ decays}
The phase space function for the interference term can be found
in~\cite{holik}
\beq
F_{ps}^{int}(x)=1+9x-9x^2-x^3+6x(1+x)\ln(x).
\eeq

\secb{\bll}
\beqa
G^{SM}(x)&=&\frac{1}{2} \lef(
x+\frac{3x}{1-x}+\frac{3x^2\ln(x)}{{(1-x)}^2} \rit), \\
J(y_t)&=&\frac{1-y+\ln(y)}{2{(1-y)}^2}, \cont
I_9(y_t)&=& J_2(1,1,y_t,0)=
\frac{1-y_t-y_t\ln(y_t)}{{(1-y_t)}^2},   \cont
I_{10}(x_t,x_H)&=&J_2(1,x_t,x_H,0) =
\frac{x_t\ln(x_t)}{(1-x_t)(x_H-x_t)} +
\frac{x_H\ln(x_H)}{(1-x_H)(x_t-x_H)}. \nonumber
\eeqa


\begin{thebibliography}{99}
\bibitem{hunter}{J.F. Gunion, H.E. Haber,
 G.L. Kane and S. Dawson, {\it The Higgs Hunter's Guide}
 (Addison-Wesley Publishing Company, Reading, MA, 1990)
 and references therein.}
\bibitem{pdg}
 {K. Hikasa et al. (Particle Data Group), Phys. Rev. D45 (1992) S1.}
\bibitem{glwein}
 {S.L. Glashow and S. Weinberg, Phys. Rev. D15 (1977) 1958.}
\bibitem{tye} {C. Albright, J. Smith and S.-H.H. Tye,
 Phys. Rev. D21 (1980) 711.}
\bibitem{bbg}
 {G.C. Branco, A.J. Buras and J.-M. Gerard, Nucl. Phys. B259 (1985) 306.}
\bibitem{wein}{S. Weinberg, Phys. Rev. Lett. 37 (1976) 657.}
\bibitem{branco}
 {G.C. Branco, Phys. Rev. lett. 44 (1980) 504.}
\bibitem{pok2}
 {P. Krawczyk and S. Pokorski, Nucl. Phys. B364 (1991) 10.}
\bibitem{yuval} {Y. Grossman and Y. Nir, Phys. Lett. B313 (1993) 126.}
\bibitem{qmass}{J. Gasser and H.Leutwyler, Phys. Rep. 87 (1982) 77.}
\bibitem{cdf}{D. Amidei, CDF Collaboration, talk given in
 the fifth international symposium on heavy flavor physics, Montreal,
 Canada (1993).}
\bibitem{lep}{A. Blondel and C.Verzegnassi,
 Phys. Lett B311 (1993) 346.}
\bibitem{barger2}
 {V. Barger and R.J.N. Phillips, Phys. Rev. D41 (1990) 884.}
\bibitem{guth}
 {A. Denner, R.J. Guth, J.H. K\"{u}hn, Phys. Lett. B240 (1990) 438.}
\bibitem{yckm}
 {Y. Nir, Lectures presented in TASI-91, SLAC-PUB-5676 (1991)
 and references therein.}
\bibitem{nirsarid}{Y. Nir and U. Sarid, Phys. Rev. D47 (1993) 2818
 and references therein.}
\bibitem{ship}
 {J. Bartelt {\it et al.}, CLEO Collaboration,
  Phys. Rev. lett. 71 (1993) 4111.}
\bibitem{buras}
 {A.J. Buras, P. Krawczyk, M.E. Lautenbacher and C. Salazar,
 Nucl. Phys. B337 (1990) 284.}
\bibitem{ron}
 {A. Sopczak, CERN-PPE/93-86.}
\bibitem{barger}
 {V. Barger, J.L. Hewett and R.J.N. Phillips,
 Phys. Rev. D41 (1990) 3421.}
\bibitem{fran}{P.J. Franzini, Phys. Rep. 173 (1989) 1.}
\bibitem{ycpv}
 {Y. Nir, Lectures presented in the 20th SLAC Summer Institute,
 SLAC-PUB-5874 (1992) and references therein.}
\bibitem{sach}{C.T. Sachrajda, Nucl. Phys. B. Proc. Supp. 30 (1992) 20.}
\bibitem{jamin}{A.J. Buras, M. Jamin and P.H. Weisz,
 Nucl. Phys. B347 (1990) 491.}
\bibitem{bbhls}{G. Buchalla, A.J. Buras, M.K. Harlander,
 M.E. Lautenbacher and C. Salazar,
 Nucl. Phys. B355 (1991) 305.}
\bibitem{wise}
 {L.F. Abbott, P. Sikivie and M.B. Wise, Phys. Rev. D21 (1980) 1393.}
\bibitem{gil}{G.G. Athanasiu, P.J. Franzini and F.J. Gilman,
 Phys. Rev. D32 (1985) 3010.}
\bibitem{mati}{M. Neubert, Phys. Rev. D45 (1992) 2451.}
\bibitem{gil2}{G.G. Athanasiu and F.J. Gilman,
 Phys. Lett. B153 (1985) 274.}
\bibitem{cheng}{H.-Y. Cheng, Int. J. Mod. Phys. A7 (1992) 1059.}
\bibitem{dupont}{Y. Dupont and T.N. Pham, Phys. Rev. D28 (1983) 2169; \\
 H.Y. Cheng, Phys. Rev. D34 (1986) 1397; \\
 J.S. Hagelin, Phys. Lett. B117 (1982) 441; \\
 H.Y. Cheng, Phys. Rev. D42 (1990) 2329.}
\bibitem{ddbigi}{I.I. Bigi and A.I. Sanda, Phys. Lett. B171 (1986) 320;
 \\ I.I. Bigi, UND-HEP-89-BIG01 (1989).}
\bibitem{woldon}{L. Wolfenstein, Phys. Lett. B164 (1985) 170; \\
 J.F. Donoghue {\it et al.}, Phys. Rev. D33 (1986) 179.}
\bibitem{georgi}
 {H. Georgi, Phys. Lett. B297 (1992) 353; \\
  T. Ohl, G. Ricciardi and E.H. Simmons, Nucl. Phys B403 (1993) 605.}
\bibitem{altar}
 {G. Altarelli, R. Barbieri and F. Caravaglios, CERN-TH.6859/93.}
\bibitem{park}
 {G.T. Park, CTP-TAMU-69/93.}
\bibitem{finnell}
 {M. Boulware and D. Finnell, Phys. Rev. D44 (1991) 2054.}
\bibitem{dghk}
 {A. Denner, R.J. Guth, W. Hollik and J.H. K\"{u}hn,
 Z. Phys. C51 (1991) 695.}
\bibitem{cleo}{E. Thorndike, CLEO Collaboration, talk given in
 the meeting of the American Physical Society, Washington D.C. (1993).}
\bibitem{misiak}
 {A.J. Buras, M. Misiak, M. M\"{u}nz and S. Pokorski,
 MPI-Ph/93-77 (1993).}
\bibitem{lim}{T. Inami and C.S. Lim, Prog. Theor. Phys. 65 (1981) 297;
 (E) 65 (1981) 1772.}
\bibitem{hou}{W.-S. Hou and R.S. Willey,
 Phys. Lett. B202 (1988) 591; Nucl. Phys. B326 (1989) 54.}
\bibitem{rizzo}{T.G. Rizzo, Phys. Rev. D38 (1988) 820.}
\bibitem{gsw}{B. Grinstein, R. Springer and M. Wise,
 Nucl. Phys. B339 (1990) 269.}
\bibitem{ma}{E. Ma and A. Parmudita, Phys. Rev D24 (1981) 1410.}
\bibitem{e787}{M.S. Atiya {\it et al.}, E787 collaboration,
 Phys. Rev. Lett. 70 (1993) 2521.}
\bibitem{bigi}{I.I. Bigi and F. Gabbiani, Nucl. Phys. B367 (1991) 3.}
\bibitem{geng}{C.Q. Geng and J.N. Ng, Phys. Rev. D38 (1988) 2858;
 (E) 41 (1990) 1715.}
\bibitem{buch1}{G. Buchalla and A.J. Buras, Nucl. Phys B398 (1993) 285.}
\bibitem{aleph}{D. Buskulic {\it et al.}, ALEPH collaboration,
 Phys. Lett. B298 (1993) 479; \\
 A. Putzer, ALEPH Collaboration, talk given in
 the fifth international symposium on heavy flavor physics, Montreal,
 Canada (1993).}
\bibitem{flnn}
 {A.F. Falk, Z. Ligeti, M. Neubert and Y.Nir, WIS-93/117/Dec-PH.}
\bibitem{pokor}
 {P. Krawczyk and S. Pokorski, Phys. Rev. Lett. 60 (1988) 182.}
\bibitem{kalino}{J. Kalinowski, Phys. Lett. B245 (1990) 201; \\
 B. Grzadkowski and W.-S. Hou, Phys. Lett. B272 (1991) 383.}
\bibitem{isidori}{H. Isidori, Phys. Lett. B298 (1993) 409.}
\bibitem{wstro}{A.J. Weinstein and R. Stroynowski, CALT-68-1853 (1993).}
\bibitem{haber}{H.E. Haber, G.L. Kane and T. Sterling,
 Nucl. Phys. B161 (1979) 493.}
\bibitem{holik}{W. Holik and T. Sack, Phys. Lett. B284 (1992) 427.}
\bibitem{roney}
 {M. Roney, talk given in
 the fifth international symposium on heavy flavor physics, Montreal,
 Canada (1993).}
\bibitem{beall}
 {G. Beall and N.G. Deshpande, Phys. Lett. B132 (1983) 427.}
\bibitem{dicus}{D.A. Dicus, Phys. Rev. D41 (1990) 999.}
\bibitem{ckly}{D. Chang, W.Y. Keung, C.S. Li and T.C. Yuan,
 Phys Lett. B241 (1990) 589.}
\bibitem{buch2}{G. Buchalla and A.J. Buras, Nucl. Phys B400 (1993) 225.}
\bibitem{hnr}
 {J.L. Hewett, S. Nandi and T.G. Rizzo, Phys. Rev. D39 (1989) 250.}
\bibitem{he}
 {X.-G. He, T.D. Nguyen and R.R. Volkas, Phys. Rev. D38 (1988) 814.}
\bibitem{skiba}
 {W. Skiba and J. Kalinowski, Nucl. Phys. B404 (1993) 3.}
\bibitem{desh}{N.G. Deshpande, K. Panose and J. Trampetic,
 Phys. Lett. B308 (1993) 322.}
\bibitem{cleoblnu}
 {D. Cinabro, CLEO Collaboration,
 talk given in the Fermilab Meeting DPF 92 (1992).}
\bibitem{hou2}{W.-S. Hou, Phys. Rev. D48 (1993) 2342.}
\bibitem{hewett}
 {J.L. Hewett, Phys. Rev. Lett. 70 (1993) 1045; \\
 V. Barger, M.S. Berger and R.J.N. Phillips,
 Phys. Rev. Lett. 70 (1993) 1368.}
\bibitem{bigipri}{A. Pich, CERN-TH.7066/93.}
\bibitem{eilam}
 {D. Atwood, G. Eilam, A. Sony, R.R. Mendel and R. Migneron,
  Phys. Rev. Lett. 70 (1993) 1364.}
\bibitem{cortes}
 {J.L. Cortes, X.Y. Pham and A. Tounsi, Phys. Rev. D25 (1982) 188.}
\end{thebibliography}
\end{document}